\documentclass[journal=ancac3,manuscript=article,layout=twocolumn]{achemso}

\usepackage{chemformula} 
\usepackage[T1]{fontenc} 
\usepackage[abs]{overpic}
\usepackage{bm}
\usepackage{amsmath}
\usepackage[normalem]{ulem}

\author{Gaia \@ Ciampalini}
\affiliation{Dipartimento di Fisica “E. Fermi,” Università di Pisa, Largo B. Pontecorvo 3, I-56127 Pisa, Italy}
\alsoaffiliation{Graphene Labs, Istituto Italiano di Tecnologia, Via Morego 30, I-16163 Genova, Italy}
\alsoaffiliation{NEST, CNR—Istituto Nanoscienze and Scuola Normale Superiore, piazza San Silvestro 12, I-56127 Pisa, Italy}
\email{gaia.ciampalini@phd.unipi.it}

\author{Filippo \@ Fabbri}
\affiliation{NEST, CNR—Istituto Nanoscienze and Scuola Normale Superiore, piazza San Silvestro 12, I-56127 Pisa, Italy}

\author{Guido \@ Menichetti}
\affiliation{Graphene Labs, Istituto Italiano di Tecnologia, Via Morego 30, I-16163 Genova, Italy}
\alsoaffiliation{Dipartimento di Fisica “E. Fermi,” Università di Pisa, Largo B. Pontecorvo 3, I-56127 Pisa, Italy}

\author{Luca \@ Buoni}
\affiliation{Dipartimento di Fisica “E. Fermi,” Università di Pisa, Largo B. Pontecorvo 3, I-56127 Pisa, Italy}

\author{Simona \@ Pace}
\affiliation{Graphene Labs, Istituto Italiano di Tecnologia, Via Morego 30, I-16163 Genova, Italy}
\alsoaffiliation{Center for Nanotechnology Innovation @NEST, Istituto Italiano di Tecnologia, Piazza San Silvestro 12, I-56127 Pisa, Italy}

\author{Vaidotas \@ Mi\v{s}eikis}
\affiliation{Graphene Labs, Istituto Italiano di Tecnologia, Via Morego 30, I-16163 Genova, Italy}
\alsoaffiliation{Center for Nanotechnology Innovation @NEST, Istituto Italiano di Tecnologia, Piazza San Silvestro 12, I-56127 Pisa, Italy}

\author{Alessandro \@ Pitanti}
\affiliation{NEST, CNR—Istituto Nanoscienze and Scuola Normale Superiore, piazza San Silvestro 12, I-56127 Pisa, Italy}

\author{Dario \@ Pisignano}
\affiliation{Dipartimento di Fisica “E. Fermi,” Università di Pisa, Largo B. Pontecorvo 3, I-56127 Pisa, Italy}
\alsoaffiliation{NEST, CNR—Istituto Nanoscienze and Scuola Normale Superiore, piazza San Silvestro 12, I-56127 Pisa, Italy}

\author{Camilla \@ Coletti}
\affiliation{Graphene Labs, Istituto Italiano di Tecnologia, Via Morego 30, I-16163 Genova, Italy}
\alsoaffiliation{Center for Nanotechnology Innovation @NEST, Istituto Italiano di Tecnologia, Piazza San Silvestro 12, I-56127 Pisa, Italy}

\author{Alessandro \@ Tredicucci}
\affiliation{Dipartimento di Fisica “E. Fermi,” Università di Pisa, Largo B. Pontecorvo 3, I-56127 Pisa, Italy}
\alsoaffiliation{NEST, CNR—Istituto Nanoscienze and Scuola Normale Superiore, piazza San Silvestro 12, I-56127 Pisa, Italy}

\author{Stefano \@ Roddaro}
\affiliation{Dipartimento di Fisica “E. Fermi,” Università di Pisa, Largo B. Pontecorvo 3, I-56127 Pisa, Italy}
\alsoaffiliation{NEST, CNR—Istituto Nanoscienze and Scuola Normale Superiore, piazza San Silvestro 12, I-56127 Pisa, Italy}

\title[]
 {Unexpected Electron Transport Suppression in a Heterostructured Graphene-MoS$_2$ Multiple Field-Effect Transistor Architecture}


\begin{document}

%
%

\begin{abstract}
  We demonstrate a graphene-MoS$_2$ architecture integrating multiple field-effect transistors (FETs) and we independently probe and correlate the conducting properties of van der Waals coupled graphene-MoS$_2$ contacts with the ones of the MoS$_2$ channels.
  Devices are fabricated starting from high-quality single-crystal monolayers grown by chemical vapor deposition.
  The heterojunction was investigated by scanning Raman and photoluminescence spectroscopies.
  Moreover, transconductance curves of MoS$_2$ are compared with the current-voltage characteristics of graphene contact stripes, revealing a significant suppression of transport on the $n$-side of the transconductance curve.
  Based on \textit{ab Initio} modeling, the effect is understood in terms of trapping by sulfur vacancies, which counter-intuitively depends on the field-effect, even though the graphene contact layer is positioned between the backgate and the MoS$_2$ channel. 
\end{abstract}

\section{}
Van der Waals (vdW) heterostructures, formed when two or more atomically-thin crystals are bonded by vdW interaction~\cite{geim2013van}, are intriguing architectures, enabled by the discovery of two-dimensional (2D) materials, such as graphene, hexagonal boron nitride, and transition metal dichalcogenides (TMDs).
Within this family, peculiar junctions can be obtained when graphene is used as a contact material for a TMD monolayer.
While the interface between a TMD and a conventional, bulk metallic electrode tends to display Schottky behavior due to intrinsic and extrinsic Fermi pinning phenomena~\cite{Kim:2017up,sotthewes2019universal,das2013high}, vdW graphene-TMD junctions yield well-behaved linear transport characteristics~\cite{Liu:2015wv}.
This contacting approach has been successful in improving TMD-based devices~\cite{Chuang:2014td} and, together with the side-contacting approach~\cite{L.:2013tl, Matsuda:2010tk, Yu:2021vj}, is commonly used for the realization of most devices based on two-dimensional materials.
Graphene-TMDs heterostructures were employed for many other applications, \textit{e.g.} in flexible photodetectors \cite{De-Fazio:2016vw}.
Nevertheless, the exact physics behind graphene-TMD vdW junctions is still debated~\cite{Yu:2013wp,Kwak:2014tf,Tian:2014wm} and difficult to probe in a direct way.
In particular, devices typically include only two contacts, which makes effects on the transport characteristics due to the interface not easy to distinguish from those due to the resistivity of the two-dimensional materials, since typically only the global conductance of the device can be measured.
Charge transfer phenomena~\cite{Dappe:2020wx}, strain~\cite{Liu:2015wh}, and charge trapping in defects~\cite{Han:2015uf,Piccinini:2019wy,Avsar:2014tc}, might also play an important role.
Furthermore, in the case of field-effect devices, the low density of states in the vicinity of the Dirac point leads to weak screening properties despite the metallic nature of graphene~\cite{Li:2018tr}.
This implies that a non-trivial response to field effect can be observed, and exploited in device concepts~\cite{Zhu:2019vv,Krishnaprasad:2019te}.
Gating on graphene-MoS$_2$ heterostructures has been widely investigated from a numerical~\cite{stradi2017field} and experimental~\cite{Cui:2015uk,Liu:2015wv,Lee:2015ve,Bertolazzi:2013vu} point of view, and in different stacking configurations.
Indeed, the reciprocal electrostatic screening of the junction materials can affect the contact resistance, and it was shown theoretically that MoS$_2$ can screen the field effect on graphene, or not, depending on the order of the specific stacking sequence~\cite{stradi2017field}.
Nevertheless, to the best of our knowledge, direct experimental evidence of how the formation of vdW interfaces leads to changes in the trasport properties of the individual materials involved in FET devices is still missing.

\section{Results and discussion}

\begin{figure}[t]
	\includegraphics[width=\columnwidth]{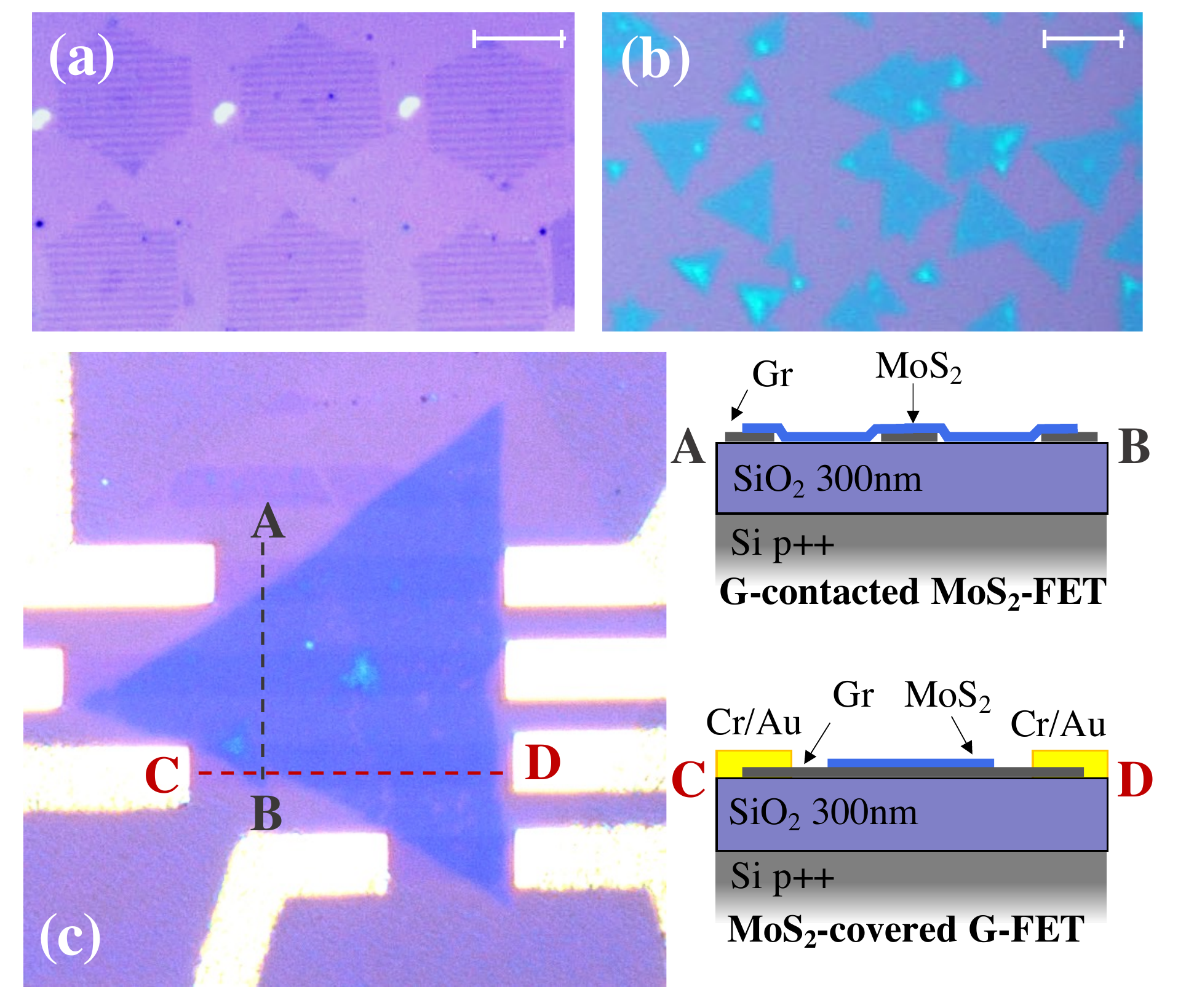}
	\caption{{\bf Multi-FET device architecture.} (a) Monocrystalline contact stripes obtained by patterning a periodic array of graphene CVD flakes (scale bar is $100\,{\rm \mu m}$). (b) Monocrystalline CVD MoS$_2$ flakes before transfer onto the SiO$_2$/Si substrate (scale bar is $50\,{\rm \mu m}$). (c) Optical picture of one of the studied devices implementing a multiple FET structure, as visible in the cross-section sketches: graphene multi-contact MoS$_2$ FET (AB section) and MoS$_2$-covered graphene FET (CD section).}
	\label{geometry}
\end{figure} 
The progress of large-scale chemical vapor deposition (CVD) techniques gives the opportunity to investigate vdW interfaces from a different angle.
High-quality and large-scale monocrystalline flakes of graphene~\cite{Giambra:2021uo,Wang:2021tm} and TMDs~\cite{Schram:2017um,Cun:2019uz} can be reproducibly grown.
When this technique is associated with a patterning of the seed points, predictable flake arrays of chosen sizes can be achieved~\cite{Miseikis:2017aa}, enabling the fabrication of multiple parallel devices combining different 2D materials.
Here, we take advantage of this opportunity to demonstrate a graphene-dichalcogenide architecture where a monocrystalline MoS$_2$ channel is contacted by a large number of monocrystalline graphene stripes, each of them crossing the whole MoS$_2$ channel as schematized in Fig.~\ref{geometry}.
Each stripe can thus act as ohmic contact for a MoS$_2$ backgated FET (see cross-section AB in Fig.~\ref{geometry}), and be simultaneously contacted at its terminations to implement an additional MoS$_2$-covered graphene FET (cross-section CD in Fig.~\ref{geometry}).
This structure can so act as a MoS$_2$ FET and as a set of graphene FETs at the same time, which will be referred to as a {\em multi-FET} in the following.
\begin{figure*}[htbp]
	\includegraphics[width=\textwidth]{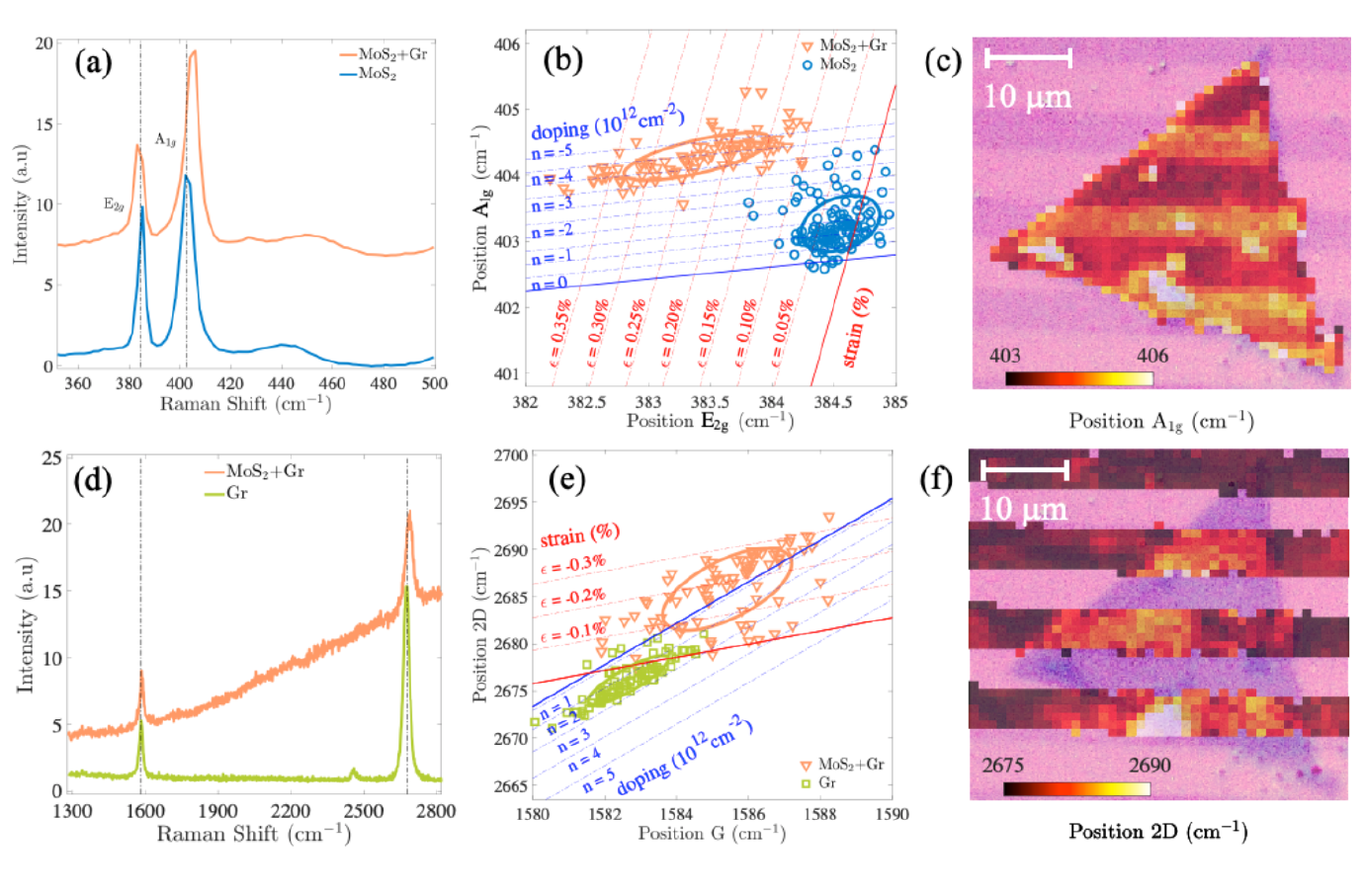}
	\caption{ {\bf Raman measurements of the MoS$_2$-graphene structures.} (a) MoS$_2$ Raman spectra after transfer on top of the graphene stripes: both spectra from MoS$_2$ on top of graphene (orange) and graphene-free MoS$_2$ (blue) are reported. (b) Correlation plot of the position of $A_{\rm 1g}$ as a function of the position of $E_{\rm 2g}$. Zero strain and zero doping line are taken from Ref.~\citenum{Michail:2016tg} (514.5 nm laser excitation).(c) Map of the position of $A_{\rm 1g}$. (d) Raman spectra of graphene after the MoS$_2$ transfer: both spectra in the presence (orange) and absence (green) of the MoS$_2$ overlayer are reported. (e) Correlation plot of the position of $2D$ peak as a function of the position of $G$ peak. Zero strain and zero doping line are taken from Ref.~\citenum{Lee:2012uk} (514.5 nm laser excitation) (f) Map of the position of $2D$ peak. Correlation plots in panel b and e were obtained from Raman spectra collected as far as possible from the flakes boundaries to avoid spill-over effects from neighboring regions and do not derive from the datasets used in panel c and f.}
	\label{Raman}
\end{figure*} 
Such arrangement allows studying the conducting properties for the different components of graphene-MoS$_2$ systems, highlighting a suppression of the electron-side transconductance in MoS$_2$-covered graphene, in coincidence with the conducting threshold of free MoS$_2$ in the main device channel.
This behavior is apparently at odd with recent predictions for defect-free MoS$_2$~\cite{stradi2017field}, nonetheless it can be understood in terms of a gate-driven trapping by sulfur vacancies which further highlights the non-trivial weak screening properties of graphene.

\begin{figure*}[htbp]
	\includegraphics[width=\textwidth]{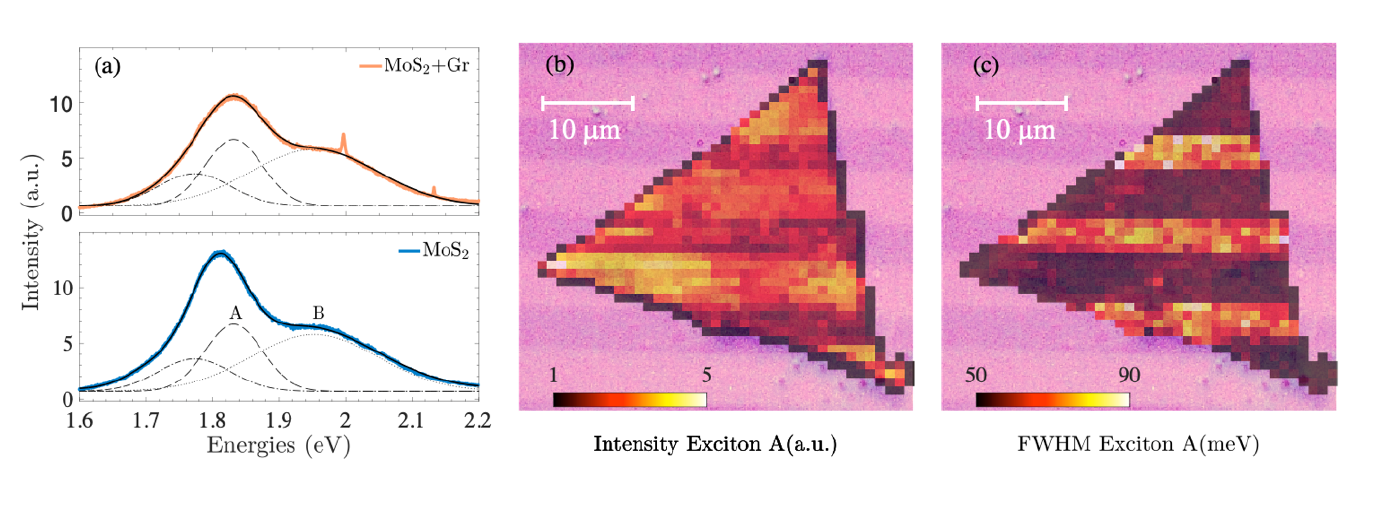}
	\caption{ {\bf Photoluminescence measurements of the MoS$_2$-graphene structures.} (a) PL spectra of MoS$_2$ both in a region where it overlaps graphene (orange) and in a graphene-free region (blue). Gaussian fits of $A$ and $B$ excitons are shown in dashed and dot-dashed lines respectively. (b) Map of the position-dependent quenching of the $A$ exciton signal. (c) Map of the position-dependent $A$ exciton broadening. All maps are shown in overlay to an optical image of the analyzed flake, scalebars in the panels correspond to $10\,{\rm \mu m}$.}
	\label{PL}
\end{figure*} 

Both the patterning procedures and the formation of vdW interfaces can significantly perturb the properties of the 2D materials.
For this reason, photoluminescence (PL) and Raman spectroscopy were employed to characterize the 2D crystals at relevant device processing steps. 
In Fig.~\ref{Raman}a-c, we report Raman spectra of a typical transferred MoS$_2$ flake and analyze the influence of the graphene contact stripes.
We analyse a MoS$_2$ flake with a small bilayer island both in regions with and without graphene. The Raman spectra of bilayer MoS$_2$-graphene heterostructure and bare MoS$_2$ bilayer are reported in Supporting Information.
The characteristic $A_{\rm 1g}$ and $E_{\rm 2g}$ modes visible in panel (a) exhibit a strong dependence on thickness~\cite{Lee:2010vo} and their separation $\Delta \omega \simeq$ 19 cm$^{-1}$ is in good agreement with the expected monolayer nature of the MoS$_2$ flake.
The shift of the A$_{1g}$ mode in Fig.~\ref{Raman} b-c can be interpreted as caused by the interlayer interaction between MoS$_2$ and graphene, as reported in Ref.~\citenum{Zhou:2014wv}. Nevertheless, a recent work~\cite{Rao:2019tu} propose an alternative interpretation in terms of doping and strain, extending the scope of a method which is typically used for bare graphene~\cite{Lee:2012uk} and bare MoS$_2$~\cite{Michail:2016tg}.
The following discussion is based on this last interpretation.
Starting from strain and doping reference lines reported in Ref.~\citenum{Michail:2016tg}, we consider the correlation plot in Fig.~\ref{Raman}b, where the position of the $A_{\rm 1g}$ peak is plotted against the one of $E_{\rm 2g}$~\cite{Michail:2016tg}.
Mean Raman shifts of the $E_{\rm 2g}$ and $A_{\rm 1g}$ peaks in graphene-free regions are $384.5\pm0.3$ cm$^{-1}$ and $403.3\pm0.4$ cm$^{-1}$, respectively.
These values are quite close to the neutrality point, located at the intersection between zero strain and zero doping line ($E_{\rm 2g}=384.6\pm0.2\,{\rm cm^{-1}}$ and $A_{\rm 1g}=402.7\pm0.2\,{\rm cm^{-1}}$)~\cite{Michail:2016tg}.
Interestingly, Raman shifts from regions where MoS$_2$ overlaps graphene ($E_{\rm 2g}=383.4\pm0.5\,{\rm cm^{-1}}$ and $A_{\rm 1g}=404.3\pm0.3\,{\rm cm^{-1}}$) indicate a variation in both tensile strain distribution of $\approx 0.10-0.35\%$ and a sizable electron reduction of $ (3.0\pm 1.8)\times 10^{12}$ cm$^{-2}$.
Thanks to the sensitivity of the $A_{\rm 1g}$ peak on doping~\cite{Chakraborty:2012tk}, the spatial doping modulation of MoS$_2$ due to graphene can be directly appreciated in the map of the $A_{\rm 1g}$ position in Fig.~\ref{Raman}c.
The map is shown in overlay to an optical picture of the flake, to highlight the good correlation between the map patterns and the position of the graphene stripes.
Consistent evidences are obtained from graphene Raman data shown in Fig.~\ref{Raman}d-f.
In panel (d), graphene spectra in the presence/absence of MoS$_2$ are compared.
Both curves show a single sharp Lorentzian-shaped $2D$ peak, which is a typical signature of monolayer graphene~\cite{Ferrari:2006wz}, and no $D$ peak, which indicates a negligible density of defects~\cite{Cancado:2011wg}.
The absence of defects was confirmed for all the fabrication steps (see Supporting Information).
We also note that when graphene is covered in MoS$_2$ (orange curve), a strong baseline appears below the Raman peaks due to the MoS$_2$ PL signal.
As in the case of MoS$_2$, strain and doping profiles can be derived from the Raman data~\cite{Lee:2012uk}, based on the correlation plot of the $2D$ and $G$ modes reported in Fig.~\ref{Raman}e, {where the strain and doping reference lines are taken from Ref.~\citenum{Lee:2012uk}, see Supporting Information for additional correlation plots.
The positions of the $G$ and $2D$ peaks in regions free from MoS$_2$ are $1582.7\pm0.9\,{\rm cm^{-1}}$ and $2676.2\pm2.2\,{\rm cm^{-1}}$, respectively, corresponding to a $p$-type doping.
Differently, Raman data collected in MoS$_2$-covered regions ($1585.4\pm1.5\,{\rm cm^{-1}}$ and $2685.8\pm3.7\,{\rm cm^{-1}}$ for the $G$ and $2D$ peaks, respectively) fall on the strain line, thus indicating a neutralization of graphene.
The variation of mean peaks positions corresponds to an electron increase of $ (2.3\pm 1.5)\times 10^{12}$ cm$^{-2}$ and to a variation of strain nature from tensile to compressive.
The spatial modulation of the doping can be seen from the $2D$ peak position map in panel (f), showing a good correlation with the position of the MoS$_2$ flake.
A modified strain is also observed, turning from slightly tensile to compressive $\approx 0.10-0.30\%$.
We note that the doping and strain trends observed where MoS$_2$ and graphene overlap are opposite and thus consistent.
We further highlight that, while the discussed analysis of the Raman data indicates an electron transfer from MoS$_2$ to graphene~\cite{Rao:2019tu}, absolute equilibrium carrier densities are not obvious to quantify. This due to the presence of photo-excited carriers during the Raman measurements, and to the unknown exact calibration of the zero-strain and zero-doping points (standard values from Ref.~\citenum{Lee:2012uk} and~\citenum{Michail:2016tg} were used).
The formation of the heterojunction can be further investigated based on the PL spectra of MoS$_2$, which are reported in Fig.~\ref{PL}a.
Three main peaks are highlighted by Gaussian deconvolution.
These peaks are attributed to the A exciton ($1.81\,{\rm eV}$), the B exciton ($1.94\,{\rm eV}$) and the trion ($1.71\,{\rm eV}$)\cite{Mouri:2013vd,Splendiani:2010ta}.
We observe that the presence of graphene modifies the MoS$_2$ response and the signal of the $A$ exciton is quenched when MoS$_2$ is coupled to graphene (orange curve) with respect to stand-alone MoS$_2$ (blue curve): indeed, a lowering of the $A$ intensity by $\approx 30\%$ and a lineshape broadening from $\approx66\,{\rm meV}$ to $\approx96\,{\rm meV}$ is retrieved.
The spatial modulation of the effect can be directly appreciated from the maps of the intensity and width of the $A$ exciton in Fig.~\ref{PL}b and Fig.~\ref{PL}c, respectively.
Additional spectroscopic data are reported in the Supporting Information.

\begin{figure}[ht!]
	\includegraphics[width=0.5\textwidth]{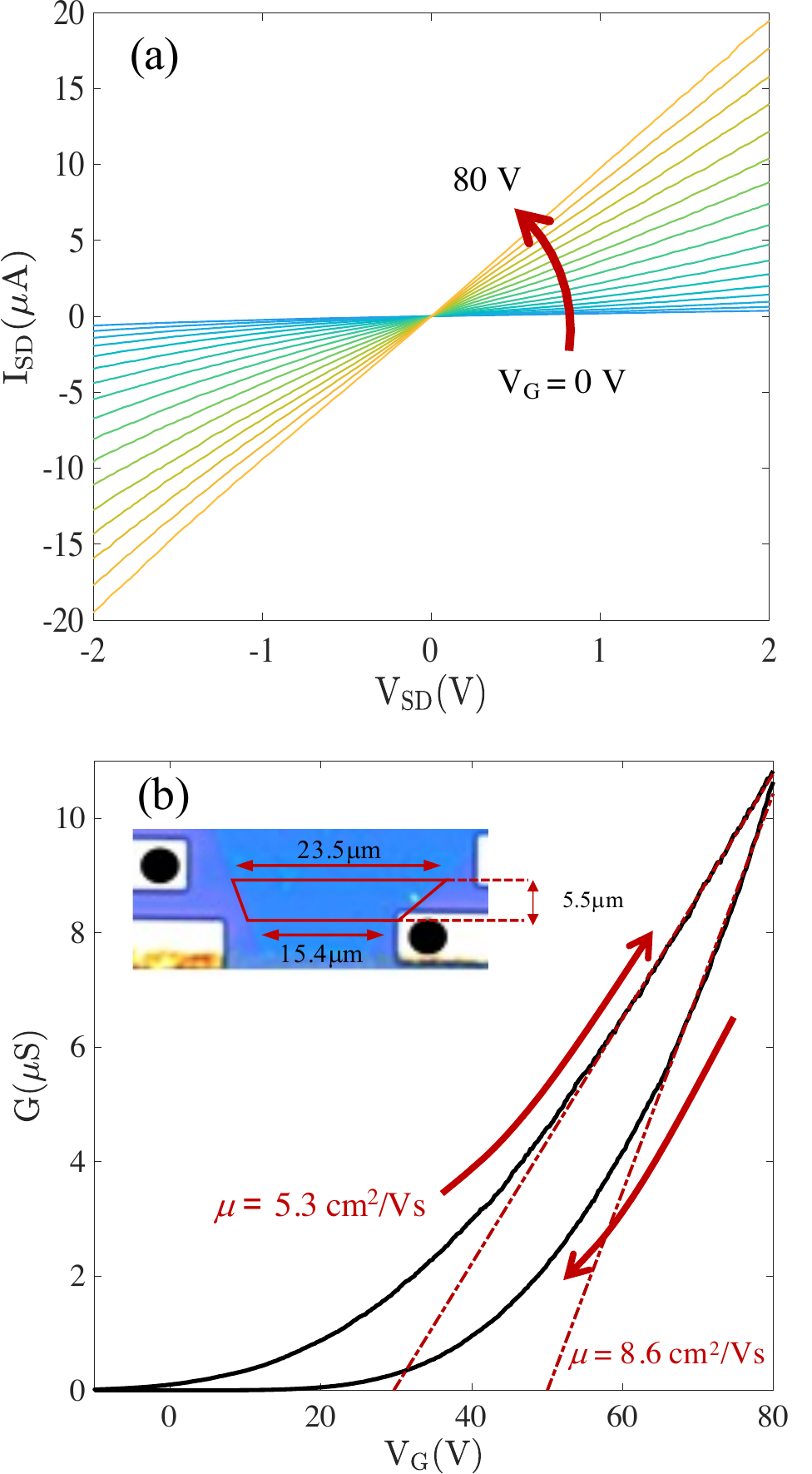}
	\caption{{\bf Transport characteristics of MoS$_2$}. (a) Room-temperature IV characteristics of the MoS$_2$ channel as function of the gate voltage $V_{\rm G}$ in the $0-80\,{\rm V}$ range. (b) Transfer characteristics showing a strong hysteresis, with red arrows indicating the sweep direction. Red dashed lines are the linear fits used to estimate the field-effect mobility for each of the two curves. Inset: an optical image of the measured device, with a sketch of the channel geometry and contacts highlighted by black dots.}
	\label{MoS2_Transport}
\end{figure} 

\begin{figure*}[ht]
	\includegraphics[width=\textwidth]{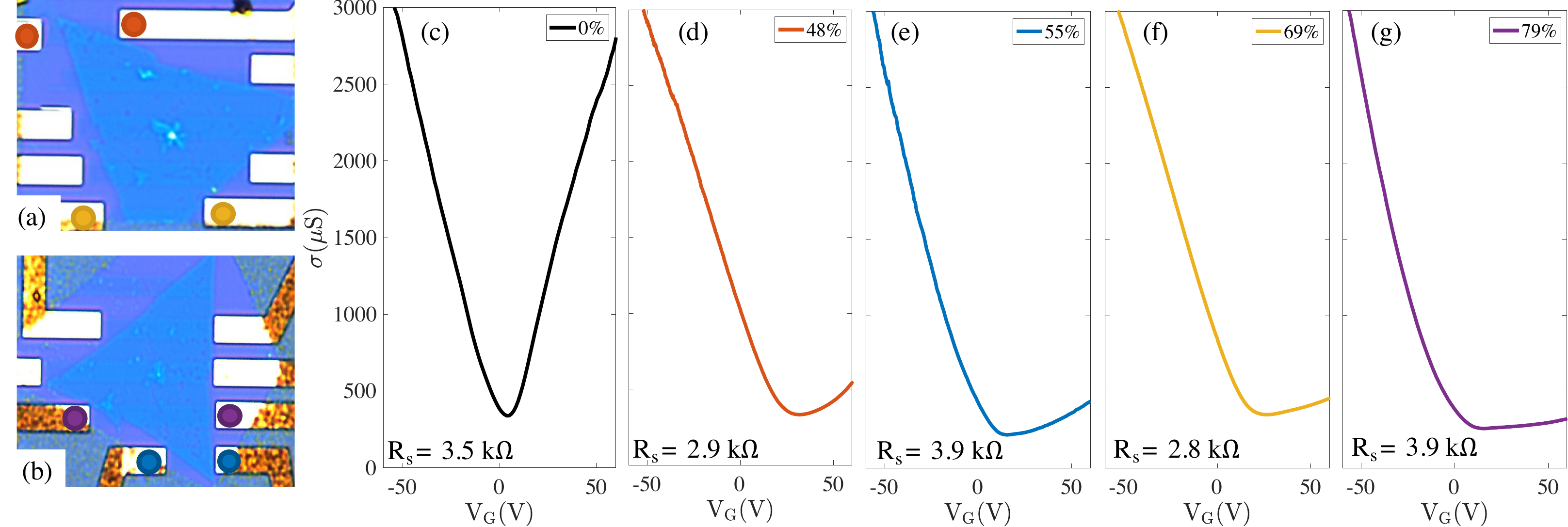}
	\caption{{\bf Effect of the MoS$_2$ overlayer on electron transport in the graphene contact stripes}. (a-b) Optical images of the two devices used to estimate the effect of different MoS$_2$ coverage levels on conduction in the graphene stripes. Used contacts are highlighted by colored dots. (c-g) Transfer characteristics of the graphene stripes for different MoS$_{2}$ coverages ranging from $0\%$ to $79\%$. The curve colors match the ones used to highlight the contacts in panels (a) and (b): red $48\%$, blue $55\%$, yellow $69\%$, purple $79\%$; the black curve corresponds to a reference MoS$_2$-free graphene stripe (device image not shown).}
	\label{Trans_G}
\end{figure*} 

The graphene stripes form ohmic contacts to the MoS$_2$ channel and lead, at room temperature and in vacuum conditions ($P<10^{-5}\,{\rm mbar}$), to highly linear two-wire IV curves over the $\pm 2\,{\rm V}$ range.
In Fig.~\ref{MoS2_Transport}a, we report the $I_{\rm SD}$ \textit{vs.} $V_{\rm SD}$ characteristics of a representative MoS$_2$ FETs, measured as a function of the gate voltage ($V_{\rm G}$) in the $0-80\,{\rm V}$ range.
The transfer characteristic in Fig.~\ref{MoS2_Transport}b indicates a positive threshold voltage, with a sizable clock-wise hysteresis, as frequently reported in literature for FETs based on 2D materials and nanowires~\cite{Mitta:2020vm, Roddaro:2013tz, Di-Bartolomeo:2017tz,Late:2012wo,Vu:2018tv}, as well as in Kelvin probe microscopy experiments~\cite{Wang:2020vq}.
The effect is generally ascribed to the slow dynamics of trap states leading to a time-dependent screening of the field effect of the gate.
Trap states may have several origins, including defects at the SiO$_2$ substrate interface~\cite{Vu:2018tv}, adsorbates~\cite{Late:2012wo}, or MoS$_2$ point defects~\cite{Di-Bartolomeo:2017tz}.
In our devices, possible sources of traps include interfaces between MoS$_{2}$, graphene and SiO$_2$ (see AFM data in the Supporting Information) as also S vacancies in MoS$_2$, which are know to occur in quite large densities  (typically few $10^{13}\,{\rm cm^{-2}}$) in CVD flakes~\cite{Hong:2015vp, Qiu:2013vd}.
The field-effect mobility of MoS$_2$ carrier can be estimated from the transfer characteristic according to
\begin{equation}
	\mu = \frac{d G}{dV_G} \frac{L^2}{C_G},
\end{equation}
where $C_G$ is the capacitance and the MoS$_2$ trapezoid channel sketched in the inset of Fig.~\ref{MoS2_Transport}b is approximated as a rectangle with a length $L = 5.5\pm 0.3\,{\rm \mu m}$ and width $W=19.5\pm0.5\,{\rm \mu m}$.
Considering both curves in the hysteresis loop, we extract two mobility values $\approx 5.3\,{\rm cm^{2}/Vs}$ and $\approx 8.6\,{\rm cm^{2}/Vs}$.
Similar analysis on different FETs yielded field-effect mobilities in the range $5.3-6.6\,{\rm cm^2/Vs}$.
Given that gate hysteresis generally indicates that part of the gate-induced carriers end in charge traps, field-effect measurements are known to overestimate the carrier density induced in the channel and to underestimate mobility~\cite{Roddaro:2013tz,Pitanti:2012vj}.
Both the mobility values above should thus be considered as a lower bound to the true room temperature electron mobility in the specific MoS$_2$ flake.
The method also neglects the effect of contact resistances, which may lead to a mobility underestimation but are not expected to have a significant effect in the explored transport regime, based on preliminary four-wire measurement data. 

Our multi-FET devices were specifically designed for comparing the MoS$_2$ transport characteristics with the electron configuration in the graphene stripes, which play here the dual role of the contact in the MoS$_2$ FET and of the channel in MoS$_2$-covered graphene FETs.
The IVs of all our graphene stripes are found to be highly linear (a representative measurement is reported in Supporting Information) and in Fig.~\ref{Trans_G} we report the transfer characteristic of various graphene FETs as a function of the $V_{\rm G}$, from which we obtain a mobility $\approx 4085\,{\rm cm^{2}/Vs}$.
These measurements are carried out at a fixed V$_{SD} = 0.2$ V, as a function of V$_G$   . 
In the plot sequence (c-g), we compare the conductivity of stripes characterized by a different MoS$_2$ coverage: conductivity is calculated from the total resistance using the geometrical form factor of the stripe and contact resistances are estimated by comparing the $p$-side of the gate sweeps; MoS$_2$ coverage is quantified from the ratio between the area of the MoS$_2$-graphene and the graphene regions, see optical pictures in Fig.~\ref{Trans_G}a-b.
Coverage goes from 0\% (MoS$_2$-free graphene in panel (c)) to $79\%$ (panel (g)).
A clear trend is observed in the transfer characteristics: curves go from a conventional ambipolar behavior in panel (c) to a limit of strongly quenched $n$-type conduction for the largest coverage in panel (g).
The observation of a quenching of field effect in graphene-TMD heterostructures has been reported in the literature, for both instance in WS$_2$ and MoS$_2$-graphene heterostructures, and in graphene functionalized with different materials, such as TiO$_2$ or organic molecules as well~\cite{Yan:2012ws,Li:2011wr, Kim:2012wb, Piccinini:2019wy, Avsar:2014tc,De-Fazio:2016vw}.
The important role of MoS$_2$ on electron transport suppression is clear as no deep suppression is observed in bare graphene stripes on SiO$_2$ devices. \cite{tyagi2021ultraclean}
On the other hand, the observed behavior is somewhat puzzling since, ideally, MoS$_2$ should not affect carrier density in graphene when positioned on top of back-gated graphene due to the reciprocal screening in the vdW heterostructure~\cite{stradi2017field}. 
\begin{figure*}[tp]
	\centering
	\begin{tabular}{cc}
		\begin{overpic}[unit=1mm,width=\columnwidth]{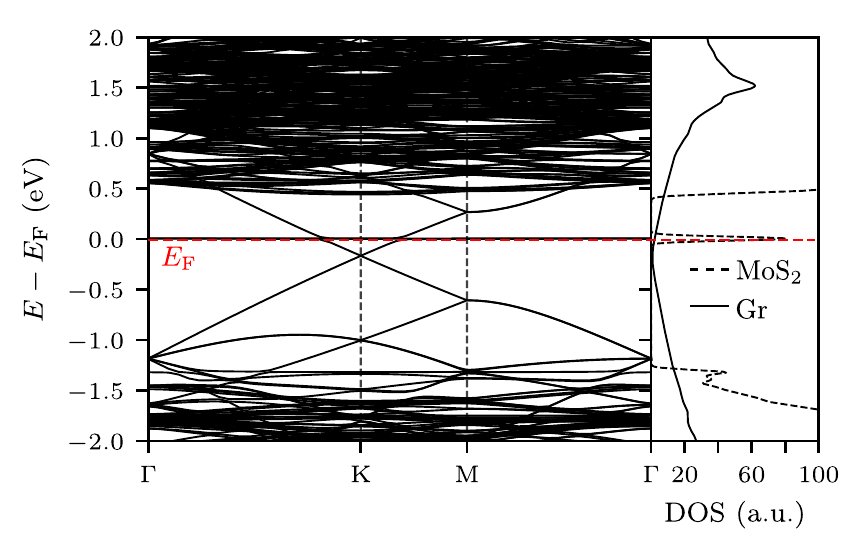} 
			\put(0,52){\rm{(a)}}
		\end{overpic} &
		\begin{overpic}[unit=1mm,width=0.9\columnwidth]{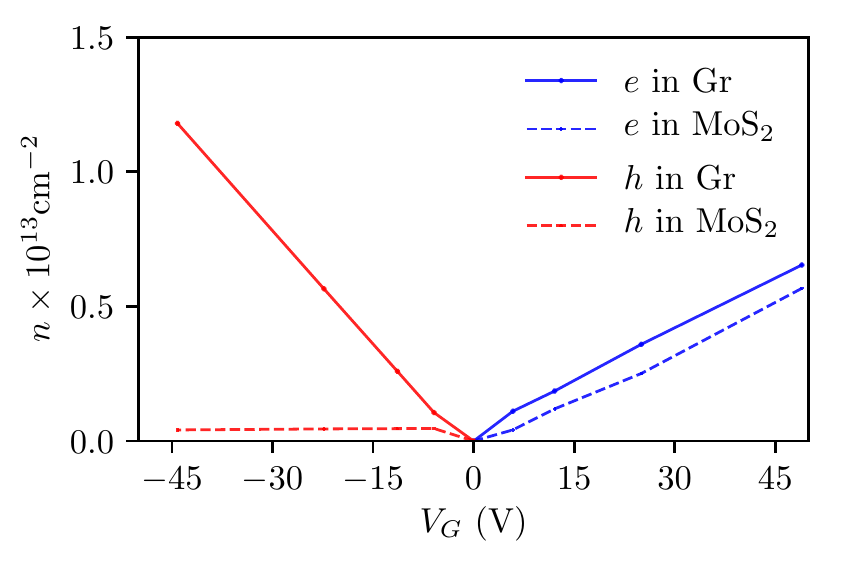}
			\put(-1,52){\rm{(b)}} 
		\end{overpic} \\
		\begin{overpic}[unit=1mm,width=\columnwidth]{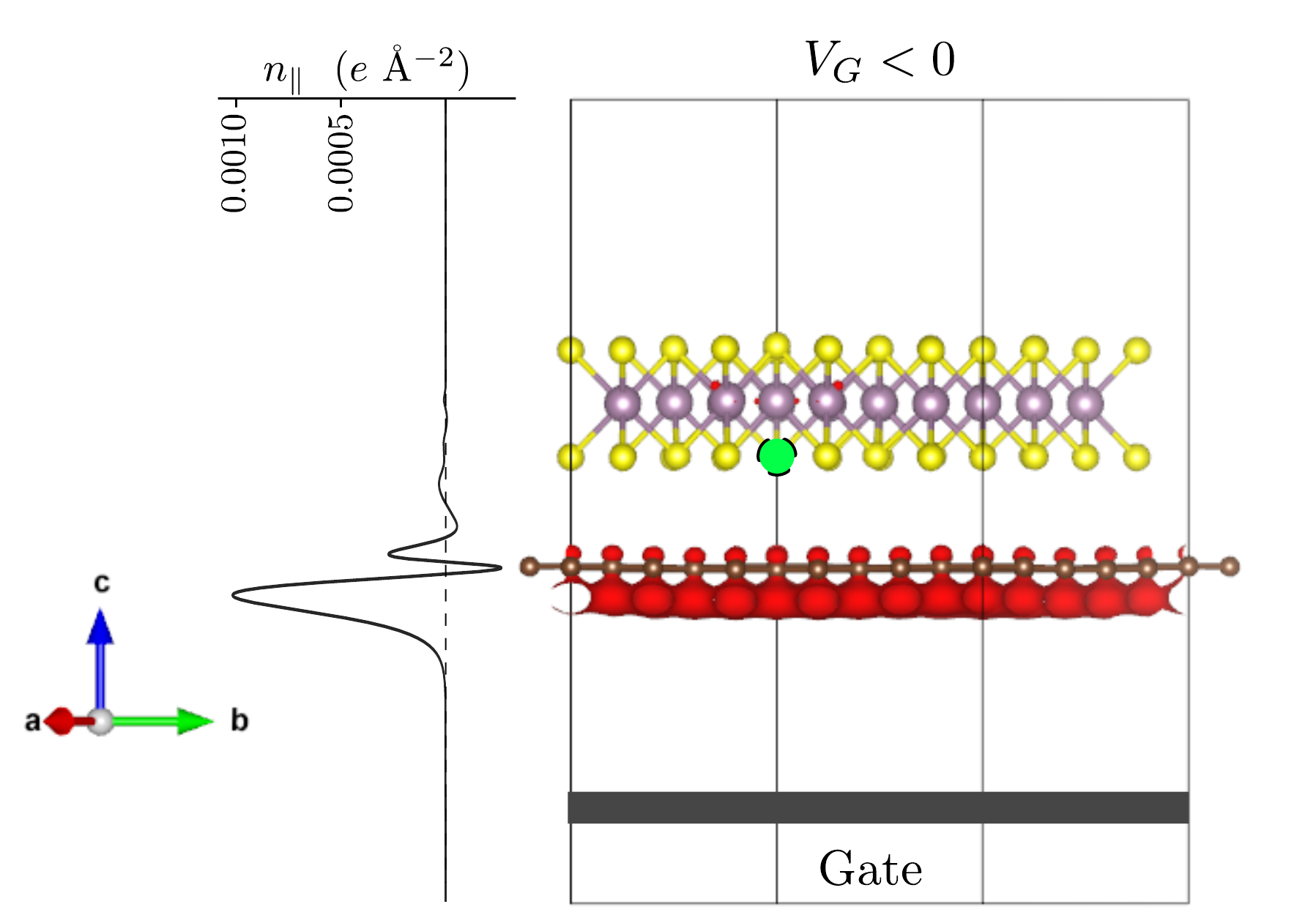} 
			\put(0,52){\rm{(c)}}
		\end{overpic} &
		\begin{overpic}[unit=1mm,width=\columnwidth]{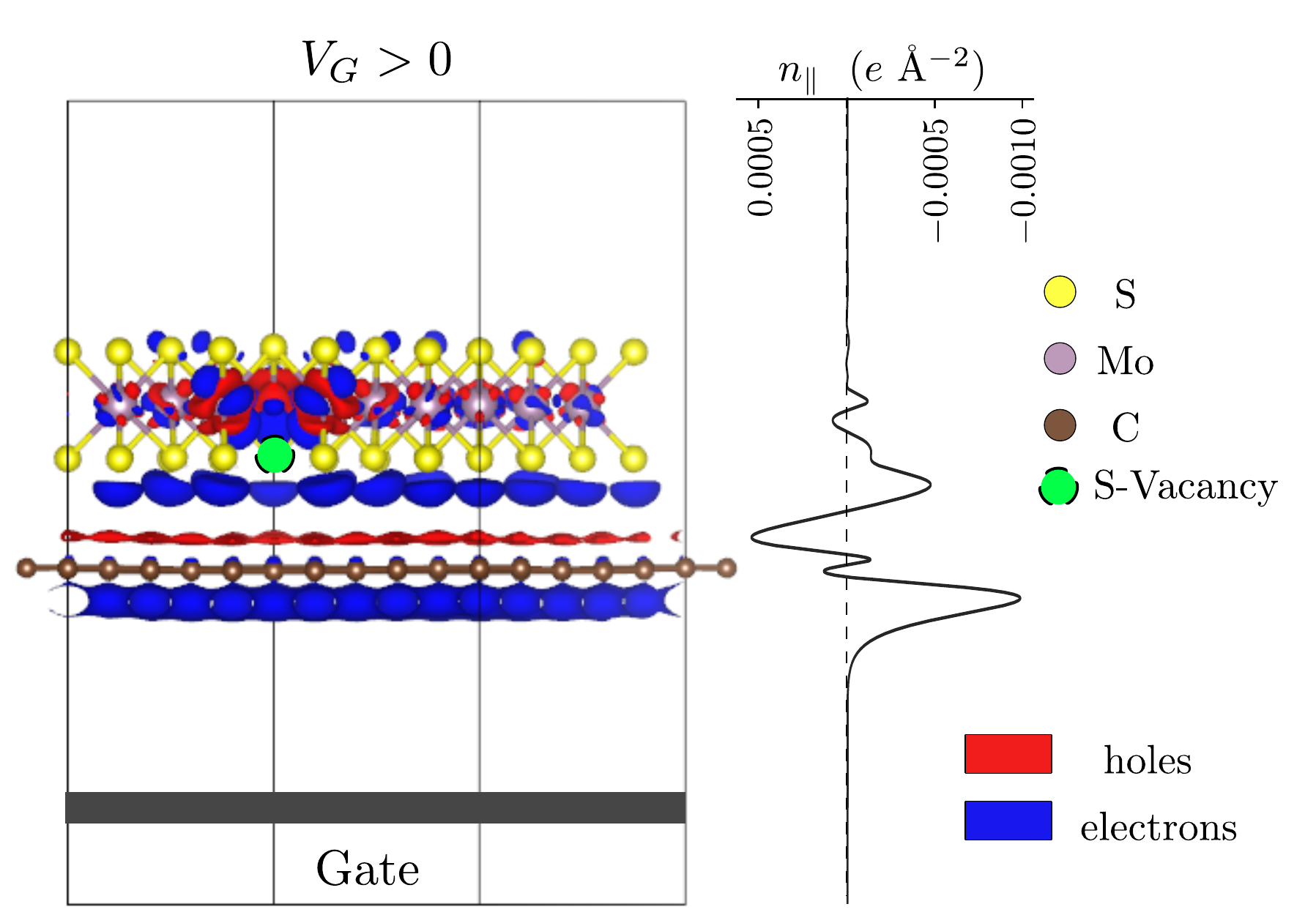}
			\put(-2,52){\rm{(d)}} 
		\end{overpic} 
	\end{tabular}
	\caption{{\bf Field-effect response in the presence of S-vacancies.} The impact of sulfur vacancies was simulated by removing one S atom from a MoS$_2$ supercell (density of S-vacancies of $\rho_v \approx1.8\times 10^{13}\,{\rm cm^{-2}}$). (a) Supercell band structure and projected density of states (DOS) of the Graphene-MoS$_2$ interface ($V_G\approx 23$ V corresponding to a charge induced by field effect $n\approx -6\times10^{12} {\rm cm^{-2}}$). The red dashed line indicates the Fermi energy $E_{\rm F}$. In the DOS plot, pristine graphene is indicated with a continuous line and the S-vacancy appears as a peak close to the Dirac point. (b) Field-effect induced charge distribution as a function of gate voltage $V_G$, evaluated as the difference between the gated ($V_G\neq 0$) and ungated case ($V_G=0$). The  solid (dashed) red  line indicates the excess holes on the graphene (MoS$_2$) monolayer, while the solid (dashed) blue  line indicates the excess electrons. (c-d) Side view of the gated Graphene-MoS$_2$ interface. In the two panels, the charge isosurface for $V_G<0$ (left) and $V_G>0$ (right) is evaluated as the difference between the charge densities for the gated and ungated limit. The location of the S-vacancy in the supercell is marked by the green ball.}
	\label{Theo}
\end{figure*}

We ascribe the origin of this apparent discrepancy as due to vacancies in TMDs~\cite{Hong:2015vp, Qiu:2013vd}.
This is a reasonable assumption, in fact, as mentioned above another possible source of trap states could be the SiO$_2$ substrate. 
Nevertheless, the effect of the SiO$_2$ substrate is secondary. In fact, the not-covered graphene stripe reported in Fig.~\ref{Trans_G}c has a standard symmetric behavior despite the presence of the SiO$_2$ substrate. Moreover, the major contribution to the electron transport suppression from the sulfur vacancies can be deduced from the behavior reported in Fig.~\ref{Trans_G}c-g where a clear trend can be observed: the suppression increases as the MoS$_2$ coverage increases.
Furthermore, it is worth noting that the electrical transport of exfoliated graphene covered by exfoliated MoS$_2$ usually does not present this suppression of the electron transport \cite{Moriya:2015wz}.
In order to corroborate our hypothesis, and highlight the effect of sulfur vacancies, we perform Density Functional Theory (DFT) calculations.

Using DFT~\cite{Grosso2014}, we made an \textit{ab Initio} analysis of the electronic states of graphene-MoS$_2$~\cite{Giuliani2005} in the presence of sulfur vacancies: these have an energy that falls in the gap~\cite{Bussolotti2021} of MoS$_2$ and are located at a distance of few Angstroms from graphene, so their effect is hard to evaluate without a first-principles approach (see Methods and Supporting Information for further details).
Numerical calculations were performed using a density of S-vacancies of $\rho_v \approx1.8\times 10^{13}\,{\rm cm^{-2}}$.
In Fig.~\ref{Theo}a we report the electronic band structure and projected density of states (DOS) of the graphene-MoS$_2$ heterostructure for $V_G=0$: as visibile in the plot, the Fermi energy of the system lays in the proximity of MoS$_2$ mid-gap states generated by the S-vacancies.
This suggests that such states may influence the mobile carrier density induced in the graphene layer by the gate when $V_G\neq 0$.
This is confirmed by calculations performed at different values of $V_G$.
As shown in Fig.~\ref{Theo}b, even if MoS$_2$ is placed on top of graphene, it affects the carrier density induced by the gate in the graphene layer.
In particular, the $n$-side of the field-effect response is reduced by $\approx 50\%$; this, combined with the likely increased scattering~\cite{Das-Sarma:2011vy,Stauber:2007tr,Qiu:2013vd} caused by the large DOS close to the Fermi energy, clearly reproduces the behavior reported in Fig.~\ref{Trans_G}.
The sulfur vacancies work as charge traps only for positive gate voltages and thus only for electron carriers.
	The sulfur vacancies generate a midgap state above the neutrality point and scattering due to midgap states reduces the conductivity as described by the formula
	
	\begin{equation}
		\sigma=\frac{2e^2}{\pi h}\frac{n_{ef\!f}}{\rho_v}ln^2(k_{\rm F} R)
	\end{equation}
	where  $n_{ef\!f}$  is the effective density of the charge carrier in the graphene sheet ($\approx~
	50\%$ of the induced doping charge), $k_F$ is the Fermi momentum due to the effective density of the charge carrier  $n_{ef\!f}$, $\rho_v$  is the density of S-vacancies, and $R=3$~\AA~is the average radius of the vacancy~\cite{Qiu:2013vd}.
	In Fig.~\ref{conductivity}, the quantitative estimation of the conductivity is shown. We used a range of $\rho_v$  similar to the one used in the \textit{ab Initio} calculations. We plot the data only for positive gate voltages because the midgap states affect the transport properties only in that sector.
In Fig.~\ref{Theo}c-d, the atomistic structure and the model of the typical setup for a field-effect measurement are shown.
The charge density isosurfaces, together with the planar averaged carrier charge density, give a pictorial view of the different behavior with negative and positive backgate voltages.
We point out that our results do not contradict, but rather complement, the conclusion drawn in recent literature~\cite{stradi2017field}: calculations also show that no charge transfer is obtained in the case of defect-free MoS$_2$ since in that limit the TMD cannot support any electron state in the relevant energy range (see Supporting Information).
Nonetheless, it is interesting to highlight that, in the presence of vacancies, a TMD overlayer can indeed have an impact on the backgate response of these vdW heterostructures and of devices based on them, extending the range of non-trivial consequences of the weak screening properties of graphene.
\begin{figure}[htbp]
	\centering
	\includegraphics[width=0.5\textwidth]{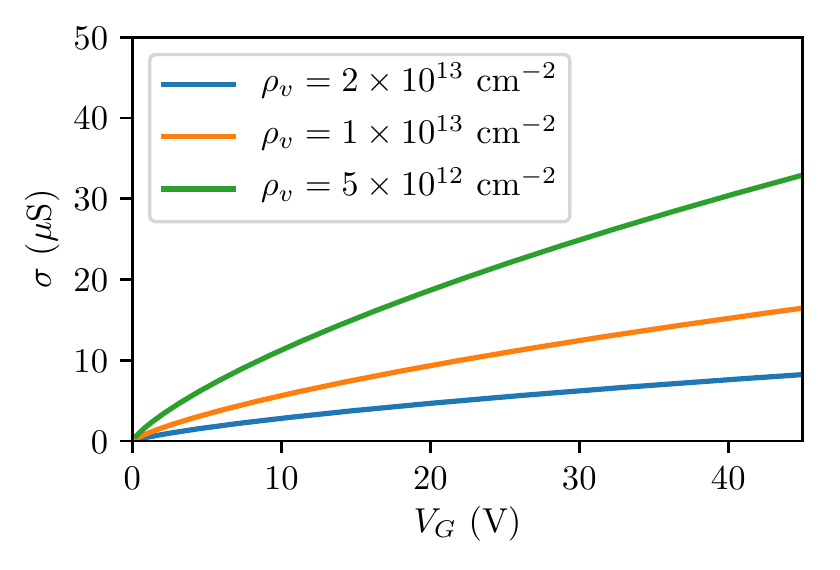}
	\caption{{\bf Quantitative estimate of the conductivity of graphene}. Midgap states associated with sulfur vacancies can suppress mobilty in graphene by increasing electron scattering. Conductivity suppression was calculated for three different densities of sulfur vacancies $\rho_v$, and using the carrier densities reported in Fig.~\ref{Theo}.}
	\label{conductivity}
\end{figure}

\section{Conclusions}
We have demonstrated a graphene-MoS$_2$ architecture integrating multiple graphene-contacted MoS$_2$ FETs and MoS$_2$-covered graphene FETs and used it to correlate the field-effect characteristics of a MoS$_2$ monolayer with the conducting properties of graphene used to contact it.
Such a study cannot be performed in a conventional FET structure since the individual resistive contributes cannot be discriminated in any obvious and direct way.
Our results show that MoS$_2$ can affect the field-effect conduction of a back-gated graphene monolayer, even when placed on top of the vdW stack, and the suppression of conduction in the graphene stripes is observed over a gate voltage range which is consistent with the conduction threshold of the MoS$_2$ channel.
This behavior is explained in terms of a filling of sulfur vacancies in the MoS$_2$, as supported by \textit{ab Initio} calculations.
The suppression of the electron transport can be exploited for the development of engineered optoelectronic devices based on van der Waals heterostructures \cite{De-Fazio:2016vw}, taking advantage of the low contact resistance of graphene in our multi-FET heterostructure. 
\section{Methods} 

{\bf Nanofabrication.}
The multi-FET fabrication starts from a square array of $\approx 150\,{\rm \mu m}$-wide single-crystal monolayer graphene flakes, with a spacing of $200\,{\rm \mu m}$.
Arrays are grown on Cu foil \textit{via} CVD~\cite{Miseikis:2017aa} and then transferred on a p++ Si substrate covered by $300\,{\rm nm}$ thermal SiO$_2$ using a delamination procedure and a semi-dry method based on a PMMA vector~\cite{Giambra:2021uo}. 
Before the MoS$_2$ transfer, the samples were cleaned from PMMA using an overnight immersion in acetone, followed by 2 minutes rinse in isopropanol, 3 minutes in AR 600-71 remover, and finally in deionized water.
The next fabrication step was the patterning of graphene into a set of $5\,{\rm\mu m}$-wide and $5\,{\rm \mu m}$-spaced stripes.
To this aim, we spun PMMA AR-P679.04 and baked the samples at $120\,{\rm ^\circ C}$ for 5 minutes.
The stripe patterns were defined \textit{via} electron-beam litography (EBL) using a \textit{SEM Zeiss Ultraplus} with a Raith lithographic module, an energy of $20\,{\rm keV}$ and a dose of $300\,{\rm \mu C /cm^2}$. The samples were then developed in AR 600-56 for 2 minutes and a half.
Then, graphene was etched by means of reactive ion etching (RIE) using Ar and O$_2$ (5:80 sccm). Finally, the samples were again cleaned from PMMA with an overnight immersion in acetone and isopropanol rinsing.
Single-crystal MoS$_2$ monolayer flakes with an average size of $\approx 50\,{\rm \mu m}$ were grown \textit{via} CVD following Ref.~\citenum{Conti:2020vh} and~\citenum{Kim:2017vz}.
Single-crystal monolayer MoS$_2$ flakes were then transferred on the graphene stripes, using a semi-dry method~\cite{Conti:2020vh, Kim:2017vz}.
The transfer process employed for MoS$_2$ is very similar to the one for graphene except for the delamination step, which was obtained by immersing the sample in a $1\,{\rm M}$ solution of NaOH rather than by an electrochemical method~\cite{Conti:2020vh}.
Given the chosen spacing between the stripes, the process typically yields various devices with $4-5$ contacts and, since the flakes are triangular, with an uneven coverage of the graphene stripes.
A final post-transfer patterning was performed to remove excess material, using a laser writer \textit{Micro Writer ML3} and a S1818 photoresist mask with a $300\,{\rm nm}$ PMMA interlayer to protect the 2D materials form contamination by the photoresist.
We then cleaned the samples with warm acetone (20 minutes) and chloroform for an hour and a half. To complete the devices we defined a set of Cr/Au ($10/50\,{\rm nm}$) metallic electrodes, \textit{via} EBL, evaporation and lift-off.
Using this method, $10$ devices were fabricated in two batches and 3 multi-FETs were measured.

{\bf Experimental Section}
The properties of graphene and MoS$_2$ were monitored by Raman and photoluminescence spectroscopy, using a \textit{Renishaw InVia} spectrometer equipped with a $532$ nm laser and 100 $\times$ objective lens (N.A 0.85).
Laser power was $\approx1\,{\rm mW}$ and the typical acquisition time was $4\,{\rm s}$~\cite{Pace:2021ut}.
Transport measurements were performed in a vacuum chamber using source-measure units K4200 and K2614B and a Femto DDPCA-300 current preamplifier.

{\bf Numerics.} We carried out DFT calculations by using the \textsc{Quantum Espresso} (\textsc{QE})~\cite{QE1,QE2,QE3,Brumme2014,Brumme2015}, which uses a plane wave basis set.
The pseudopotentials were taken from the standard solid-state pseudopotential (SSSP) accuracy library~\cite{Prandini2018,Lejaeghere2016,SG15,PSlibrary100,GBRV} with increased cutoffs of $50\,{\rm Ry}$ and $400\,{\rm Ry}$ for the wave functions and the density.
The exchange-correlation potential was treated in the ${\rm GGA}$, as parametrized by the Perdew-Burke-Ernzerhof (PBE) formula~\cite{PBE}, with the van der Waals (vdW)-D2 correction as proposed by Grimme~\cite{Grimme2006}. 
For the BZ integrations, we employed a Marzari-Vanderbilt smearing~\cite{Marzari1999} of $10^{-3}\,{\rm Ry}$ with a Monkhorst-Pack (MP)~\cite{MP} ${\bm k}$-point grid with $18\times 18\times 1$~($24\times 24\times 1$) points for self-consistent calculations of the charge density (density of states).
The heterostructure of monolayer MoS$_2$ on top of monolayer graphene is shown in Fig~\ref{Theo}c-d, where a $8\times8$ MoS$_2$ supercell is placed on a $10\times10$ supercell of graphene. The considered heterostructure model contains 391 (392) atoms in the unit cell for the simulation with (without) S-vacancy, corresponding to a   density of S-vacancies of $\rho_v \approx 1.8\times 10^{13}\,{\rm cm^{-2}}$. We keep the lattice constant of graphene unchanged at $a_0= 2.46$~\AA~\cite{Grosso2014,Colle2016} and compressed the lattice constant of MoS$_2$ by roughly $\approx 2.4\%$: from 3.15~\AA~\cite{Wakabayashi1975} to 3.075~\AA. We considered a supercell with about $18$~\AA~ of vacuum along the $\hat{\bm c}$-direction between periodic images. We optimize the geometrical structures by relaxing only the atomic positions until the components of all the forces on the ions are less than $10^{-3}~{\rm Ry}/{\rm Bohr}$, while we keep fixed the lattice parameters.

In Fig.~\ref{Theo}c-d , a model of the typical setup for a field-effect measurement is shown. The Graphene-MoS$_2$ is placed in front of a metal gate. The layers are then charged with the same amount of opposite charge,  leading to a finite electric field in the region between the heterostructure and the gate. In order to avoid spurious and artificial electric field between the different slabs of the repeated unit cell, an electric field generated by a dipole plate of opposite charge has been included next to the gate.
Furthermore, to avoid the direct interaction between the charge-density of the system and the gate, a potential barrier has been included~\cite{Brumme2014,Brumme2015}.
In order to mimic the experimental values, in Fig.~\ref{Theo}b,  we rescaled the values of $V_G$ considering that in the real experiment there is a $300\,{\rm nm}$ thick layer of SiO$_2$ between the metal gate and the Graphene-MoS$_2$ interface.
We use the \textsc{VESTA}~\cite{VESTA} code to visualize the geometrical structure, the isosurfaces, and to produce the plots in Fig.~\ref{Theo}. 
To obtain information on  the charge transfer between the two moieties (Graphene-MoS$_2$) we performed  a topological analysis of the electron density  by means of the Bader procedure~\cite{BADER,GATTI,Menichetti2016, Menichetti2017} as implemented in CRITIC2~\cite{OTERO14,OTERO9}.

\begin{acknowledgement}

G.M. thanks G. Grosso for useful discussions.
G.C. thanks L. Martini for useful discussions.
SR and AT acknowledge the support of the Italian Ministry of Research through the PRIN projects ``Quantum 2D'' and ``Monstre 2D'', respectively. G.M.  acknowledges  the ``IT center'' of the University of Pisa, the HPC center (Franklin) of the IIT of Genova, and the allocation of computer resources from CINECA, through the ``ISCRA C'' projects ``HP10C9JF51'', ``HP10CI1LTC'', ``HP10CY46PW'', ``HP10CMQ8ZK''. 
The research leading to these results has received funding from the European Union’s Horizon 2020 research and innovation program under grant agreement no. 881603-Graphene Core3.

\end{acknowledgement}

\begin{suppinfo}

Additional Raman data, additional photoluminescence data, atomic force miscroscopy measurement, additional DFT results.

\end{suppinfo}

\bibliography{GMoS2}

\end{document}